# High Performance X-ray Transmission Windows Based on Graphenic Carbon

Sebastian Huebner, Natsuki Miyakawa, Stefan Kapser, Andreas Pahlke,
and Franz Kreupl, Senior Member, IEEE

*Abstract*—A novel x-ray transmission window based on graphenic carbon has been developed with superior performance compared to beryllium transmission windows that are currently used in the field. Graphenic carbon in combination with an integrated silicon frame allows for a window design which does not use a mechanical support grid or additional light blocking layers. Compared to beryllium, the novel x-ray transmission window exhibits an improved transmission in the low energy region (0.1 keV – 3 keV) while demonstrating excellent mechanical stability, as well as light and vacuum tightness. Therefore, the newly established graphenic carbon window, can replace beryllium in x-ray transmission windows with a nontoxic and abundant material.

## I. Introduction

X-RAY transmission windows have a significant impact on the performance of x-ray applications such as vacuum encapsulated x-ray detectors and x-ray generation tubes. Besides the requirement of high x-ray transmission, the window needs to exhibit high mechanical strength and gas tightness. The low atomic order number, high mechanical strength, and gas tight nature of thin beryllium foils allow the fabrication of beryllium x-ray transmission windows that satisfy the requirements and make beryllium the most commonly used window material for medium to high x-ray energies.

However, thin beryllium foils are of limited supply and the toxicity of beryllium poses a health risk that requires safety procedures during the manufacturing and manipulation of the beryllium windows [1]. Beryllium foils are only gas tight above a minimum thickness of 8 µm, limiting the transmission for low to medium energy x-ray radiation [2]. Beryllium reacts readily with numerous chemicals including water vapor and therefore, beryllium windows require a protective coating for passivation, reducing the transmission of the window further [3]. While beryllium poses a health risk, up to now, no alternative is available without severe drawbacks. Other window materials such as high performance polymers [4], [5], diamond [6], [7], or silicon nitride [2] thin films have been proposed as alternatives. While these window materials can be made very thin and thus offer a high x-ray transmission, they lack the high mechanical strength of beryllium to support large window geometries. Therefore, a mechanical support structure is required if a significant differential pressure is applied across the window. This leads to a signal loss of more than 20 % as the support structure is radiopaque for low to medium energy x-ray radiation [5]. In addition, light blocking of the alternative window materials is insufficient for x-ray detector modules that are used in ambient light conditions. Hence, an additional metallic light blocking layer needs to be added which reduces the transmission further and introduces additional stray lines in the spectrum [2][4].

Currently, these drawbacks are tolerated for special low energy x-ray transmission windows for the detection of elements in the energy range of 0.1 keV – 1 keV as beryllium windows are not available for this low energy range, but the discussed alternatives are not a suitable replacement for beryllium windows for medium to high x-ray energies [5].

In contrast, graphite or even an individual graphite layer, called graphene, exhibits all the prerequisites for a superior window material such as low atomic order number, high mechanical strength [8], [9], gas tightness [10], high chemical stability [11], high electrical conductivity [12], light blocking ability [13], unlimited supply, and nontoxicity [14].

## II. Graphenic Carbon

The high mechanical strength of graphene is expected to allow for a reduced thickness of the x-ray transmission window without the use of a supporting structure. In a pioneering work, Bunch et al. [10] have studied monolayer graphene membranes with a diameter of 4.75 µm, which proved to be helium leak tight [10]. While the membrane itself was gas tight, gas diffusion was observed through the substrate and graphene interface region of the monolayer membranes. Hence, the interaction between the graphene layer and the substrate is of crucial interest for the fabrication of gas tight membranes.

This paragraph of the first footnote will contain the date on which you submitted your paper for review. The work has been financially supported by the Bavarian Ministry of Economic Affairs and Media Energy and Technology under contract number MST-1210-0006/ BAY 177/001.

S. Huebner and F. Kreupl are with the Institute of Hybrid Electronic Systems, Technische Universität München, Arcisstr. 21, 80333 Munich, Germany (e-mail: Sebastian.Huebner@tum.de)

N. Miyakawa, and A. Pahlke are with Ketek GmbH, Hofer Str. 3, 81737 Munich, Germany

S. Kapser was with the Institute of Hybrid Electronic Systems, Technische Universität München, Arcisstr. 21, 80333 Munich. He is now with the Max-Planck-Institut für Plasmaphysik, Boltzmannstr. 2, 85748 Garching, Germany

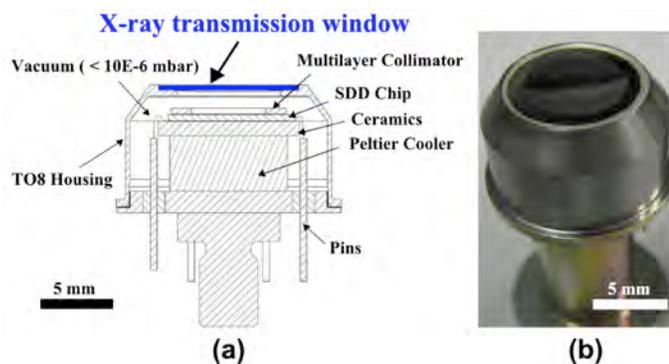

Fig. 1. (a) shows a schematic of the KETEK x-ray sensor module that is used as a testing vehicle in this work. (b) depicts a prototype detector module with a graphenic carbon transmission window integrated in the top of a TO8 housing.

In previous works, chemical vapor deposition (CVD) growth on sacrificial substrates and subsequent transfer of graphene onto the target substrate [8], [9], mechanical exfoliation [10], [12], or liquid phase deposition [11] have been used to synthesize graphene.

The simple transfer of graphene or graphite onto the target substrate results in a low adhesion interface, as the bonding graphene layer only interacts with the substrate via van der Waals forces [15]. In contrast to this, we demonstrate in this work that a direct CVD growth on oxide free silicon leads to inherently strong adhesion by silicon-carbon covalent bonding [16] and enables a gas-tight configuration.

Transmission windows that are used for x-ray detectors in an ambient light environment require high light blocking properties to avoid unwanted charge carrier generation. An attenuation of $10^{15}$ is seen as sufficient to prevent a degradation of the signal to noise ratio of the detector [2]. Graphene has a white light attenuation of 2.3 % per monolayer [13] and a specified thickness of 0.34 nm per monolayer [8]. This leads to an estimated thickness requirement of 505 nm for a graphene-multilayer transmission window to be considered light tight.

The properties of graphenic carbon (GC) [17] as an x-ray window material are evaluated in this work by fabricating GC windows and integrating them into commercially available energy dispersive spectroscopy (EDS) detector modules. Fig. 1 (a) shows the schematic of such a detector module. The silicon drift detector (SDD) is vacuum encapsulated and the TO8 housing contains a thin, highly transparent x-ray transmission window with an open diameter of 7 mm. Fig. 1 (b) shows a prototype with an integrated GC window. The silicon drift detector (SDD) is vacuum encapsulated for efficient Peltier cooling and requires a gas tight housing with an x-ray transmission window. The vacuum encapsulation of the detector module leads to a differential pressure of one atmosphere across the window if the module is used outside of a vacuum environment. The corresponding mechanical load needs to be tolerated by the transmission window.

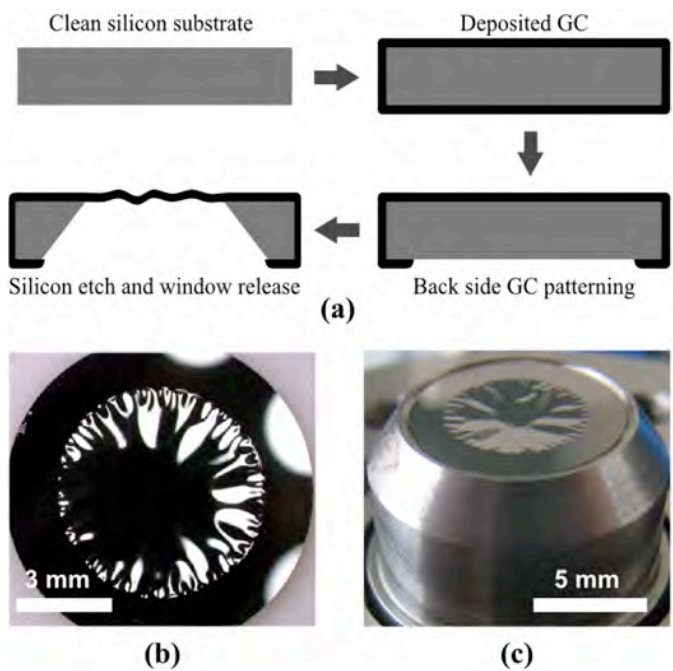

Fig. 2. (a) sketches the fabrication process of the transmission window. (b) shows a top view image of a fabricated GC window. (c) depicts a TO8 housing with a GC window glued into the top of the housing

## III. FABRICATION

In order to evaluate the GC x-ray transmission windows, round Si (100) substrates with a diameter of 9.75 mm were used for the window fabrication. The substrates were cleaned in an ultrasonic bath using acetone and isopropanol. The native oxide was removed prior to the GC deposition by a 5 % hydrofluoric acid dip. GC films with varying thicknesses from 210 nm to 2.2 μm were deposited by CVD deposition as described in [16]. The final film thickness is determined by the deposition time of the process and the slow growth rate allows excellent thickness control with an accuracy of better than 10 nm. Fig. 2 (a) illustrates the basic fabrication steps. GC is deposited directly on pre-cleaned, oxide free, Si substrates by CVD deposition. The GC on the back of the substrate is used as an etch mask for a subsequent silicon wet etch which removes the silicon in the center of the substrate. The desired window geometry is defined by structuring the GC on the backside of the Si substrate with an oxygen plasma. The patterned GC film protects the Si substrate during the hot potassium hydroxide wet etch, while the bare silicon in the center is removed. The remaining silicon substrate forms the window frame and exhibits high adhesion to the GC membrane. The addition of isopropanol to the hot potassium hydroxide etchant solution allows the fabrication of round window structures, without the use of more complex and expensive etching techniques.

Due to the compressive stress of the GC film, wrinkles, as shown in Fig. 2 (b), form in the GC thin film as soon as the Si substrate is removed. This is assumed to arise from the lower thermal expansion coefficient of graphenic carbon compared to the Si substrate [18]. As illustrated in Fig. 2 (c), the GC windows are glued into a TO8 housing in a final step before further evaluation.

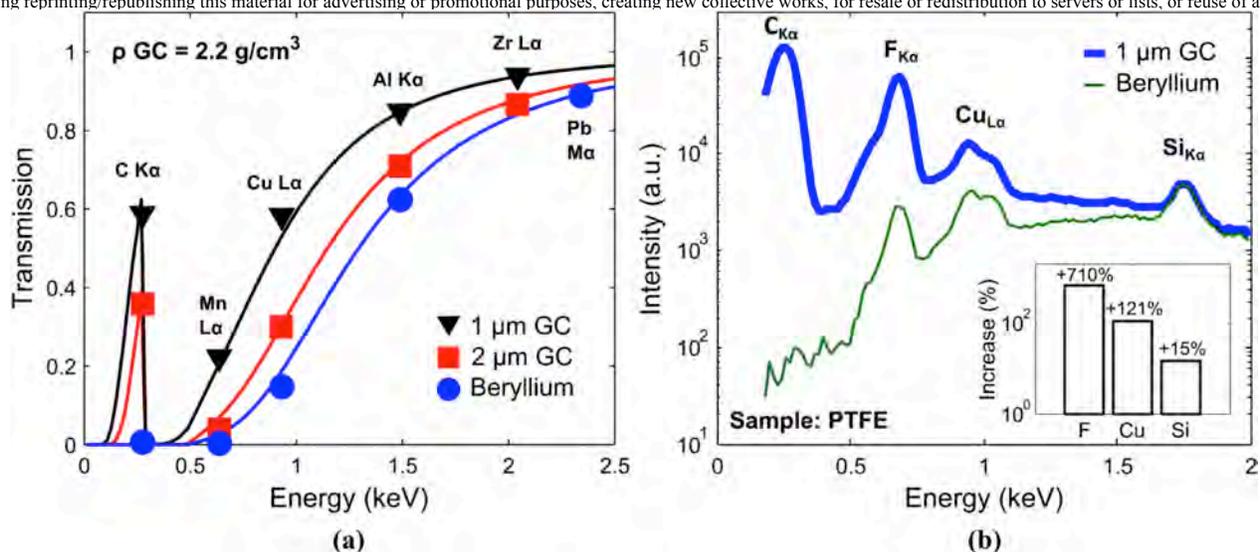

Fig. 3. (a) shows the measured x-ray transmission of beryllium [3], 1 μm, and 2 μm thick GC windows. Transmission values were experimentally obtained for discrete energies and fitted (solid line) using the transmission theory of Henke et al. [20]. Uncertainties of the data points are smaller than the symbols. (b) displays the measured EDS spectrums of PTFE as obtained with a 1 μm thick GC window and a corrosion resistant beryllium window. The inset in (b) shows the expected increase of transmission for specific elements if a 1 μm thick GC window is used instead of a corrosion resistant beryllium window.

## IV. EXPERIMENTAL

The GC windows are evaluated in regard to their x-ray transmission, mechanical stability, differential pressure cycle stability, gas tightness, and light blocking capability.

The x-ray transmission of the GC windows is determined for the energy range of 0.1 keV to 2.5 keV using an SEM/EDS system with two parallel working SDD detector modules. A focused electron beam with an acceleration voltage of 10 kV is used to excite a high purity calibration sample (BAM-EDS-TM002). One detector module serves as a reference in order to compensate for the variation of the beam current, while the second detector is used to record the counts obtained with and without a transmission window placed in front of the detector. Comparing the normalized counts of the measurements with and without the transmission window, allows the determination of the transmission factors for element specific, discrete energies. For the transmission measurements the $C_{K\alpha}$ (277.0 eV), $Mn_{L\alpha}$ (637.4 eV), $Cu_{L\alpha}$ (929.7 eV), $Al_{K\alpha}$ (1.49 keV), $Zr_{L\alpha}$ (2.04 keV), and $Pb_{M\alpha}$ (2.35 keV) lines are used [19]. The high peak to background signal of the obtained measurements allows the extraction of the transmission coefficients with an accuracy of better than 5 %.

The mechanical strength of the fabricated GC windows is determined by applying and increasing a differential pressure load until window failure. The maximum burst pressure is recorded and correlated to the GC thickness. The window thickness is determined by removing a small part of the GC that covers the window frame, and measuring the step height using an atomic force microscope. A cycling test, which periodically loads and unloads the x-ray transmission windows with a differential pressure of one atmosphere, is used to test the dynamic stability of the GC transmission windows. The pressure cycles are performed with a frequency of 0.25 Hz and reach a maximum differential pressure of 930 mbar. This results in a high stress on the x-ray transmission window along the edge of the silicon frame during the dynamic testing. Dynamic stress occurs during the venting of vacuum applications or during handling of the detector module throughout the detector lifetime.

A commercial helium leak tester (Pfeiffer HLT 570) is used to evaluate the helium leak rates of the fabricated GC windows. The detection limit of the leak tester is specified with $10^{-12}$ mbar L/s. Helium leak rates below $1 \times 10^{-10}$ mbar L/s are considered sufficient to support the high quality vacuum throughout the life time of the detector [2]. In addition, the helium leak rate is also evaluated after the GC windows have been stressed with more than 100 k pressure cycles of a differential pressure of one atmosphere in order to rule out micro crack formation during the cycle testing.

The light blocking ability of the GC window material is evaluated using a commercial Hamamatsu (C10507-11-100C) photomultiplier with a photon detection efficiency of 45 % and a white LED light source. TO8 housings with windows of 1 μm thick GC, and 2 μm thick GC are placed in front of the photomultiplier and illuminated. The photomultiplier output is used to evaluate the light blocking ability.

In order to put the obtained results into perspective, the x-ray transmission and light blocking experiments are also simultaneously conducted with beryllium windows. The used beryllium windows are specified with a thickness of 8 μm (+5 μm/-0 μm) and are passivated with an additional coating. Campbell et al. have shown that the passivation layer can be treated as an additional 2.24 μm of beryllium that is added to the nominal window thickness [3].

## V. RESULTS

The energy dependent x-ray transmission of the GC window material is plotted in Fig. 3 (a) and shows a significant improvement for the GC windows compared to the beryllium window. The measured, discrete, data points are used to generate a continuous transmission curve by fitting the material density $\rho$ to the transmission theory of Henke et al. [20]. The measured transmission of the beryllium window corresponds to a beryllium thickness of 13 µm, including the passivation layer, which is within the specified window thickness of 10.24 µm to 15.24 µm.

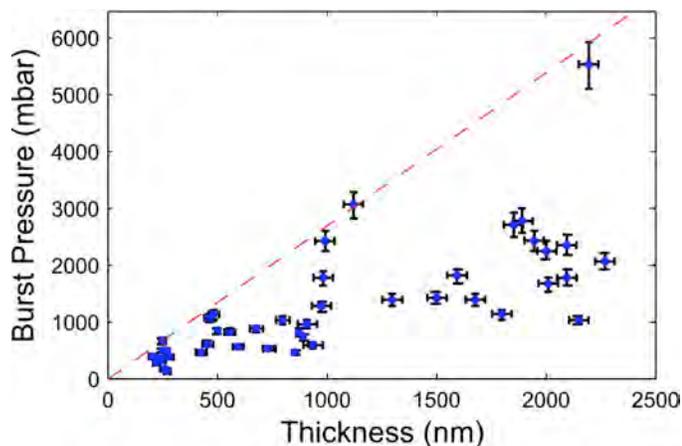

Fig. 4. Measured thickness dependent burst strength of fabricated GC transmission windows with a diameter of 7 mm.

The improved transmission of the 1 µm thick GC window allows the detection of the elements fluorine and carbon. Fig. 3 (b) shows the recorded EDS spectrums of polytetrafluoroethylene (PTFE) as obtained with a 1 µm thick GC window and a beryllium window. The copper and silicon peaks are assumed to arise from the substrate holder. The measured spectrums are normalized relative to the silicon peak at 1.74 keV for comparability. Calculations based on the transmission theory of Henke et al. [20] indicate that the transmission for radiation corresponding to the fluorine peak (676.8 eV) is increased by more than 700 % if a 1 µm thick GC window is used instead of a corrosion resistant beryllium window. In fact, the measured EDS spectrum shown in Fig. 3 (b) exhibits an increase in transmission for radiation corresponding to the fluorine peak of 2300 %, which is a result of the increased beryllium thickness of approximately 13 µm compared to the nominal thickness of 10.24 µm, used for the calculations. The low energy radiation corresponding to the carbon peak (277.0 eV) is absorbed by the beryllium window while the GC window exhibits a high transmission of 59 %, a value that is approaching those of special ultrathin low energy transmission windows [2].

Due to the vacuum environment of the used SEM system, wrinkles form in the GC transmission windows during the transmission measurements. This is in contrast to the pressurized window, as the wrinkles are stretched out if the window is used outside of a vacuum environment. Estimations conclude that the wrinkles have a minor impact on the x-ray transmission as the amplitude of the formed wrinkles is small and the radius of curvature of the GC film is very large compared to the thickness of the windows.

The mechanical strength of the fabricated windows is dependent on the window thickness as shown in Fig. 4. The maximum observed strength is plotted for reference and indicates the upper limit for the fabricated, 7 mm diameter, windows. Error bars indicate the uncertainty of the measurements. A large variation in the maximum strength of GC windows with similar thickness is observed, and most likely a direct result of defects in the deposited thin films. The given window geometry exhibits a critical area of 38.5 mm$^2$ which is large for the clean room class 1000 mini-environment used for the window fabrication. Hence for most of the tested windows, the limiting factor is assumedly not the intrinsic strength of the GC material but extrinsic defects that lead to an early failure under stress.

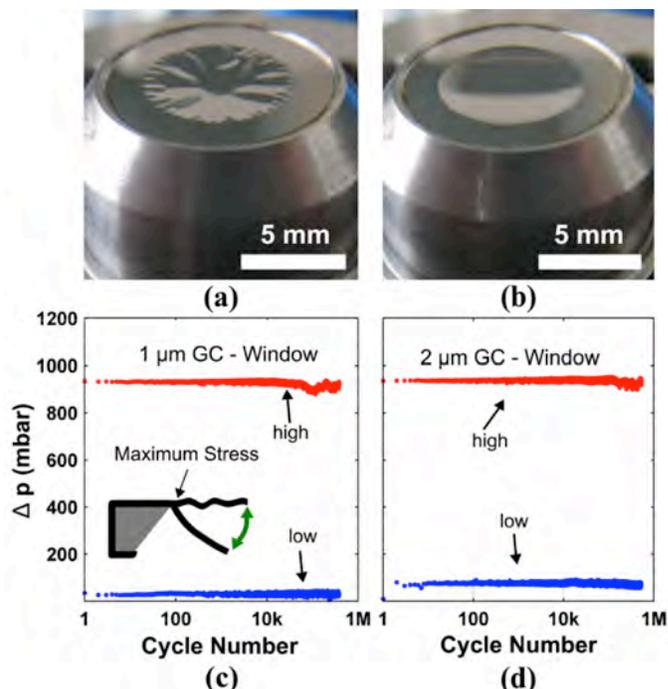

Fig. 5. (a) shows a side view optical image of a TO8 housing with GC window without applied vacuum and (b) with an applied vacuum in the housing. Both, 1 µm (c) and 2 µm (d) thick GC transmission windows were cycled for more than 500 k pressure cycles. For each pressure cycle the maximum (high) and minimal (low) differential pressure is plotted.

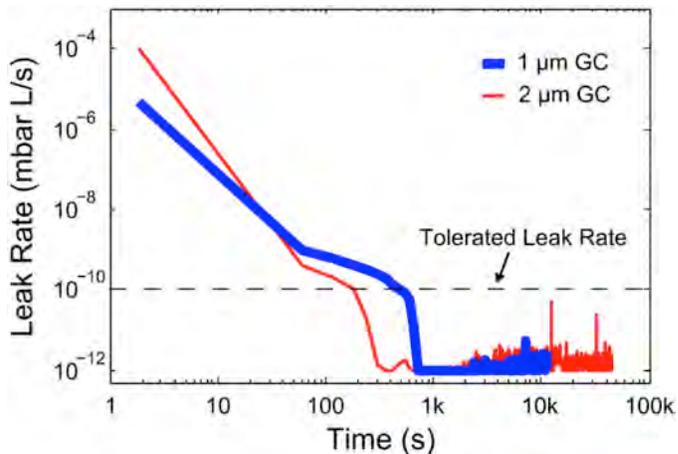

Fig. 6. Results of the helium leak tests with 1 µm thick and 2 µm thick GC transmission windows with a diameter of 7 mm. The tolerated leak rate of $1 \times 10^{-10}$ mbar L/s is indicated in the diagram.

Beryllium windows with an opening diameter of 7 mm are specified to withstand a differential pressure of 2 bars in order to guarantee the mechanical stability of the detector module. The obtained measurements show that GC windows with a thickness of 1 μm are able to exceed this requirement.

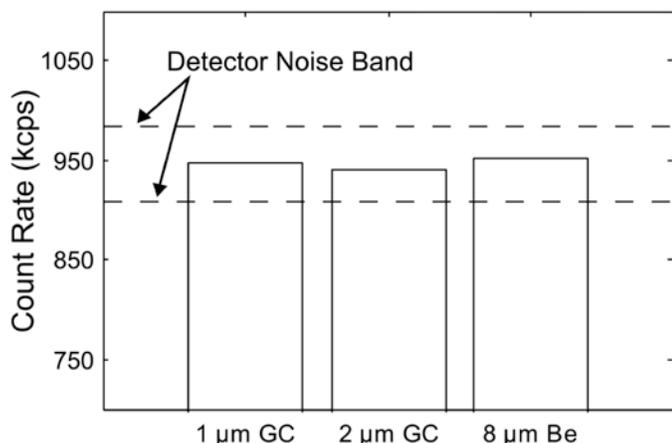

Fig. 7. Results of the illumination measurement with 1 μm and 2 μm thick GC, and 8 μm thick beryllium windows.

TABLE I:
Comparison of beryllium specifications and results of the GC window

| X-Ray Window Requirements | Beryllium | GC (This work) |
|---|---|---|
| No Support Grid at 7 mm Opening Diameter | Yes | Yes |
| X-Ray Transmission | 71 % @ 1.5 keV | 85 % @ 1.5 keV |
| Pressure Stability > 2bar | Yes | Yes |
| Pressure Cycle Fatigue | >20 k cycles | >500 k cycles |
| Helium Leak Rate | <1×10⁻¹⁰ mbar L/s | <1×10⁻¹⁰ mbar L/s |
| Thickness Tolerance | - 0 μm / +5 μm | < 10 nm |
| Light Tight | Yes | Yes |
| Chemical Resistance | High | High |
| Non-Toxic | No | Yes |
| Availability/Supply | Limited | Unlimited |

Due to the compressive stress, GC windows show wrinkles that disappear as a differential pressure is applied. This behavior is illustrated at the top of the sensor housing shown in Fig. 5 (a) and (b). The pressure cycle tests, plotted in Fig. 5 (c) and (d), showed no yielding or crack formation for as many as 500 k pressure cycles with a differential pressure of one atmosphere for 7 mm diameter windows with 2 μm and 1 μm GC thickness.

The gas tightness of the fabricated GC windows is well below the required maximum leak rate of $1 \times 10^{-10}$ mbar L/s, as shown in Fig. 6. This holds true for windows with a thickness of 1 μm and 2 μm and no notable difference was observed. The leak rates were obtained with GC windows that had previously undergone a cycle test with more than 100 k pressure cycles with a differential pressure of one atmosphere.

The light tightness was evaluated for 2 μm and 1 μm thick GC windows and compared to the results of a beryllium window as shown in Fig. 7. No count rate above the noise band of the photomultiplier was observed during the illumination with the LED light source, indicating a sufficiently high attenuation. The advantage of an inherent light blocking ability is not only the absence of a spectral contamination in the recorded spectrum due to x-ray fluorescence of the light blocking layer, but also the absence of yielding or delamination effects that can deteriorate the light blocking layer and reduce the life time of the detector module.

In order to assess the performance of the fabricated GC x-ray transmission windows, corrosion resistant beryllium windows with a nominal thickness of 8 μm and an open diameter of 7 mm are used as a benchmark. Table I summarizes the specifications and requirements of a commercially available beryllium x-ray transmission window and compares it to the obtained results of the GC windows evaluated in this work. The proposed GC transmission windows exceed the performance of beryllium windows.

## VI. CONCLUSION

The suitability of graphenic carbon as a window material for x-ray transmission windows has been demonstrated by equipping a vacuum encapsulated SDD detector module with a GC window. The obtained EDS spectrums display an increased transmission compared to standard beryllium windows and allows the examination of fluorine and carbon content in EDS applications.

A further reduction in window thickness, and hence a higher transmission in the low energy region, seems feasible but requires a better understanding of the large variations observed regarding the mechanical stability of the fabricated windows.

Experiments to identify the material properties such as Young's modulus, Poisson's ratio, and ultimate tensile strength are necessary for further optimization of the window material and window design. The use of a supporting grid would significantly reduce the required window thickness and makes the GC material a promising candidate for future ultrathin low energy transmission windows.